# Tunable Néel-Bloch magnetic twists in Fe$_3$GeTe$_2$ with van der Waals structure

*Licong Peng\*[†][1], Fehmi S. Yasin[†][1], Tae-Eon Park[2], Sung Jong Kim[2,3], Xichao Zhang[4], Takuro Nagai[5],*

*Koji Kimoto[5], Seonghoon Woo[6], Xiuzhen Yu\*[1]*

[1]RIKEN Center for Emergent Matter Science, Wako, 351-0198, Japan
[2]Center for Spintronics, Korea Institute of Science and Technology, Seoul, 02792, Korea
[3]KU-KIST Graduate School of Converging Science and Technology, Korea University, Seoul, 02841, Korea
[4]Department of Electrical and Computer Engineering, Shinshu University, Nagano, 380-8553, Japan
[5]National Institute for Materials Science, Tsukuba, Ibaraki 305-0044, Japan
[6]IBM T.J. Watson Research Center, New York 10598, USA
*\*Email*: licong.peng@riken.jp; yu_x@riken.jp
†These authors contributed equally

The advent of ferromagnetism in two-dimensional (2D) van der Waals (vdW) magnets has stimulated high interest in exploring topological magnetic textures, such as skyrmions for use in future skyrmion-based spintronic devices. To engineer skyrmions in vdW magnets by transforming Bloch-type magnetic bubbles into Néel-type skyrmions, the heterostructure of heavy metal/vdW magnetic thin film has been made to induce interfacial Dzyaloshinskii-Moriya interaction (DMI). However, the unambiguous identification of the magnetic textures inherent to vdW magnets, e.g., whether the magnetic twists (skyrmions/domain walls) are Néel- or Bloch-type, is unclear. Here we demonstrate that the Néel- or Bloch-type magnetic twists can be tuned in the vdW magnet Fe$_3$GeTe$_2$ (FGT) with/without interfacial DMI. We use an in-plane magnetic field to align the modulation wavevector *q* of the magnetizations in order to distinguish the Néel- or Bloch-type magnetic twists. We observe that *q* is perpendicular to the in-plane field in the heterostructure (Pt/oxidized-FGT/FGT/oxidized-FGT), while *q* aligns at a rotated angle with respect to the field direction in the thin plate by thinning bulk FGT. We find that the aligned domain wall twists hold fan-like modulations, coinciding qualitatively with our computational results.

Two-dimensional (2D) van der Waals (vdW) magnets provide an extraordinary platform for exploring magnetism and topological textures, such as magnetic skyrmions due to their unique layered structure and intrinsic long-range ferromagnetic order[1,2]. Skyrmions, being particle-like topological spin textures, have recently been observed in various vdW magnets [3–9]. As theoretically predicted, typical Bloch-type bubbles (magnetic twists) appear in centrosymmetric ferromagnets (FM) with dipolar interaction while Néel-type skyrmions favor in thin hetero-structural heavy metal/FM with interfacial Dzyaloshinskii-Moriya interaction (DMI)[8–12]. Several experimental observations support such theoretical predictions: Bloch-type bubbles, stabilized by the magnetic dipolar interaction in centrosymmetric FM, were claimed in the (001) thin plates (thinned from a bulk magnet) of vdW magnets $Cr_2Ge_2Te_6$[4] and $Fe_3GeTe_2$ (FGT)[5,8], while Néel-type skyrmions were reported in a heterostructure $WTe_2$/FGT with an estimated large interfacial DMI strength of 1.0 mJ m$^{-2}$ at the interface, much larger than their calculated critical DMI value of ~0.1 mJ m$^{-2}$ necessary for stabilizing Néel-type twists including skyrmions and domain walls[6]. The stability of Néel-type skyrmions is addressed also in multilayered films, such as the Pt/oxidized-FGT/FGT[9] and FGT/[Co/Pd]$_n$[7]. However, the direct identification of the Néel- or Bloch-type magnetic twists in vdW magnets remains elusive.

To identify the Néel- or Bloch-type magnetic twists, researchers[3–9,13] attempt to use Lorentz transmission electron microscopy (TEM) observations while simultaneously tilting the thin sample: the Lorentz TEM images present magnetic contrast for Bloch-type twists, but monotonic intensity (no contrast) for Néel-type twists when the tilt angle is zero. Whereas magnetic contrast arises for the Néel-type twists upon the sample tilt. The conclusion has been drawn that one may therefore use Lorentz TEM jointly with sample tilt as the evidence for identification of Néel-type twists. This is because that such Lorentz TEM observations are not sufficient to unambiguously identify the magnetic twist type because the majority of the contrast measured at a tilt arises from the sample-plane projection of the out-of-plane magnetization, and the Néel- or Bloch-type spin-rotating signature is not directly detected. One example is a demonstration of Bloch-type bubbles in a uniaxial FM $La_{1.2}Sr_{1.8}(Mn_{1-y}Ru_y)_2O_7$ with centrosymmetric crystal structure, where the DMI is absent[14]. In this case, although there is monotonic intensity at zero tilt, bubble domains can be discerned at non-zero tilt angle, indicating that the Bloch-type domain walls are too narrow to be detected by defocused Lorentz TEM observations at zero-tilt angle.

On the other hand, micromagnetic simulations[15,16] predict that Bloch-type skyrmions elongate along the direction perpendicular to the in-plane field while Néel-type skyrmions elongate along the field direction since the magnetic moments parallel to the in-plane bias field grow while the antiparallel one shrink. Previous Lorentz TEM observations have shown that in-plane bias fields align the *q*-vector of Néel-type helices perpendicular to the field direction in multilayered thin films[17], while changing that of the Bloch-type helices parallel to the field direction in the chiral magnet FeGe[18–20]. The results suggest that the external in-plane field can induce a 90°-rotation difference of *q* vectors for Néel and Bloch-type twists in DMI-stabilized systems. In addition, in-plane fields align skyrmion strings along the field direction in a thick FeGe thin plate relative to its periodicity[19–21]. Note that the alignment of Bloch-type twists in magnetic dipolar systems may show differences since there are non-negligible magnetostatic charges.

Here we aim to tune the magnetic twists and hence to identify their twisting nature in the target vdW magnet FGT with/without interfacial DMI. On the basis of the abovementioned theoretical predictions[15,16] for the in-plane magnetic field-aligned Bloch- and Néel-type skyrmions, we demonstrate changes of magnetic twists in FGT with applying in-plane fields. We first measure a series of defocused Lorentz TEM images at non-zero sample tilt in a heterostructure Pt/oxidized-FGT/FGT/oxidized-FGT (named heterostructure FGT below)/a (001) thin plate fabricated from bulk FGT by focused ion beam (FIB) (named fabricated FGT below). Skyrmions in the heterostructure FGT show monotonic intensity at zero tilt and the reversed contrasts at opposite tilting angles, respectively, while skyrmionic bubbles in the fabricated FGT show clear contrast at zero tilt. Then we use an in-plane magnetic field, obtained by changing the electric current of additional solenoid placed near the specimen plane in Lorentz TEM, to align the magnetic twists in two samples, and compare the field-induced $q$-vector reorientation with micromagnetic simulations. The real-space observations demonstrate that a weak in-plane magnetic field aligns the $q$-vector of magnetic twists perpendicular to the field direction in the FGT heterostructure, coinciding with that of Néel-type dominated twists in micromagnetic simulations. In contrast, the $q$-vector in the fabricated FGT rotates away from the in-plane field direction and the aligned twists hold fan-like modulations. On the basis of our experimental observations together with simulations, we discuss the possible mechanism of Néel-type twists forming in the FGT heterostructure, and alignments of twisted structures in two samples under in-plane magnetic fields.

In Figs. 1a-g, we show the Lorentz TEM observation of magnetic skyrmions while applying a normal magnetic field in the heterostructure FGT, where the FGT shows a layered crystalline structure with vdW gap (Fig. 1a) and is sandwiched by asymmetric layers of naturally oxidized-FGT (O-FGT) and capped by a Pt layer (Fig. 1b). The heterostructure FGT is then exfoliated and transferred on a $Si_3N_4$ membrane (Fig. 1b) as described in Ref. [9]. We generate a metastable skyrmion lattice (Fig. 1c) with 40 mT-field cooling (FC) from room temperature (RT) to 160 K (the external magnetic field is normally applied on the (001) FGT plane). Figures 1d-f show a single skyrmion at various sample tilts. At zero-degree tilt, the skyrmion does not produce any contrast in the defocused Lorentz TEM image (the defocused distance $\Delta f$ ~1 mm). The skyrmions' magnetic contrast increases with increasing tilt angle while holding the $\Delta f$ constant. At tilt angles of $\pm 20°$ (the dashed line indicates the rotation axis), the skyrmion produces elliptical-shaped contrast with half-dark and half-bright, and the contrast reverses with reversing the tilt direction. Note that this contrast is only discerned in Lorentz TEM images obtained at a non-zero tilt angle, no contrast is observed at zero tilt. The Lorentz TEM image contrast, therefore, corresponds to the in-plane projections of out-of-plane magnetizations with tilting: upwards in the center and downwards at the periphery of the skyrmions (oriented in the $\pm z$-direction at 0° sample tilt) as drawn in the schematic of Figs. 1u-1v. The magnetic induction field map of the tilted skyrmion, analyzed using the transport-of-intensity equation,[22] is shown in Fig. 1g. The closed-loop in-plane magnetic induction field is attributable to the tilted normal magnetizations projected in the in-plane, which induce the non-zero electron phase change such that the Lorentz TEM image shows visible contrast.

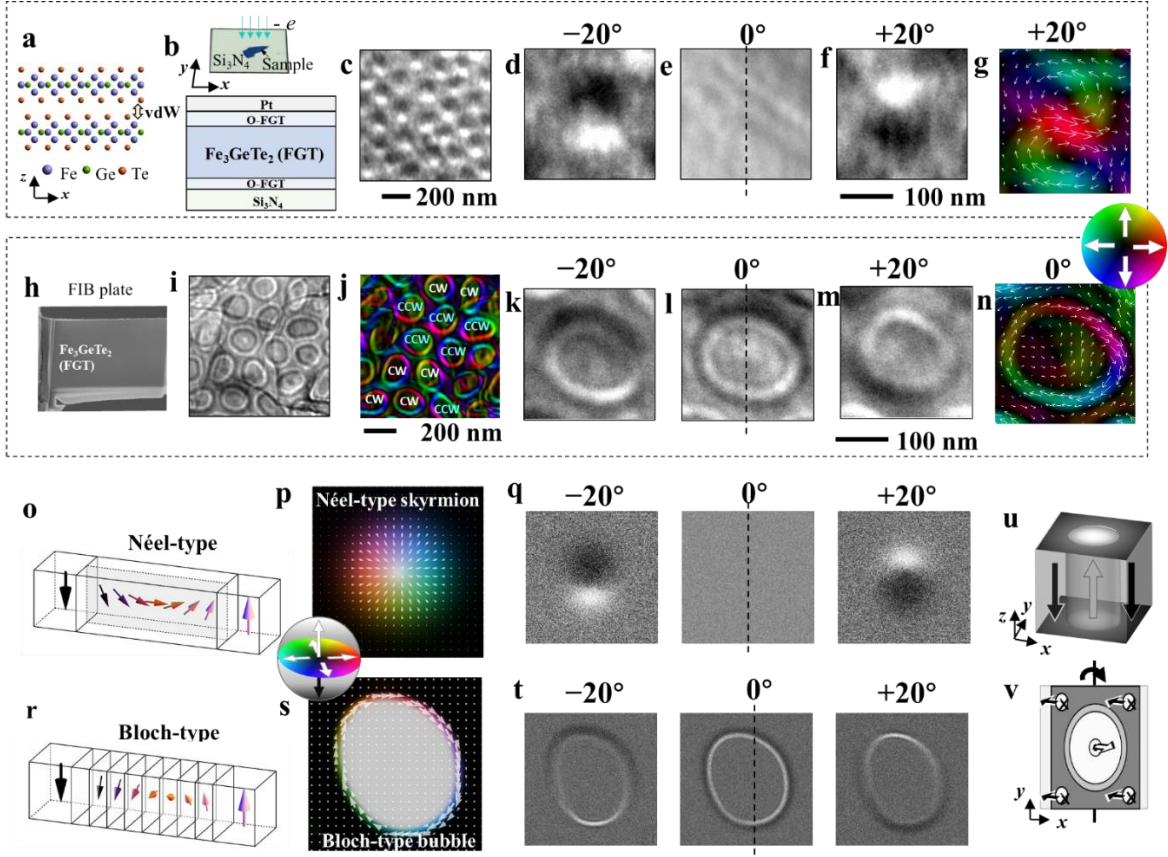

Fig 1. Magnetic skyrmionic bubbles in Fe₃GeTe₂ (FGT) magnets with van der Waals (vdW) structure. (a-g) Skyrmionic bubbles in Pt/O-FGT/FGT/O-FGT (O-FGT stands for oxidized FGT). (a) Crystal structure of FGT with vdW gap. (b) Schematic of the asymmetric multilayered heterostructure. (c-f) Lorentz transmission electron microscopy (TEM) micrographs showing (c) skyrmion lattice and (d-f) a single skyrmion with a contrast that reverses at the tilt angles of (d) −20° and (f) +20° and (e) disappears at 0°. These images were acquired after 40 mT-field cooling (FC) from room temperature (RT) to 160 K. (g) Magnetic induction field map of a skyrmion at the tilt angle of +20°. (h-n) Typical Bloch-type bubbles in a fabricated FGT thin plate. (h) Scanning electron microscopy (SEM) image of the FGT thin plate. (i) Lorentz TEM image and (j) magnetic induction field map of the magnetic bubble lattice. (k-m) Lorentz TEM images of a single bubble at the tilt angles of (k) −20°, (l) 0° and (m) +20° obtained after 150 mT - FC from RT to 100 K. (n) Magnetic induction field map of a bubble at 0° sample tilt. Schematics of (o) Néel-type twists and (r) Bloch-type twists. Multi-slice Lorentz TEM simulations of (p-q) a Néel-type skyrmion and (s-t) a Bloch-type bubble at −20°, 0°, and +20° sample tilts, respectively. (u) Schematics of a cylindrical skyrmion tube and (v) corresponding in-plane projections describing that the contrast for the tilted texture is mostly due to the in-plane projection of the tilted z-component magnetization. The color wheel indicates the direction of in-plane magnetic components, and the dark and bright contrasts indicate out-of-plane magnetic component.

Next, we measured the magnetic textures in the fabricated FGT thin plate (Fig. 1h), which has no capping layers, to compare the difference with that in the above heterostructure FGT. The Lorentz TEM images (Fig. 1i) and corresponding magnetic induction maps (Fig. 1j) show typical magnetic bubbles generated after 250 mT-FC from RT to 100 K. Two types of bubbles are presented: one, which we call skyrmionic bubbles, have a winding number $N = 1$ (as shown by the magnetic induction field map in Fig. 1n) and two helicities [*i.e.*, clockwise (CW) and counter-clockwise (CCW)] (as marked in Fig. 1j, and Fig. S1). The other exhibits $N = 0$ (see details in Fig. S1). Note that the skyrmionic bubbles are observed at zero tilt and at a relatively smaller $\Delta f$ = 200 um than that for observing skyrmions in heterostructure FGT. We conclude here that the observed skyrmionic bubbles are Bloch-type, characteristic of those stabilized by the magnetic dipolar interaction in centrosymmetric FM. We also performed a series of Lorentz TEM observations at various tilt angle, as shown in Figs. 1k-m. The magnetic contrast of the bubble is nearly symmetric at zero tilt (Fig. 1l), and show unbalanced dark/bright contrasts at $\pm 20°$ tilt that are reversed at opposite tilt angles (Fig. 1k, Fig. 1m and Fig. S2). We also attribute this asymmetric contrast to the projected in-plane components of the core and peripheral out-of-plane magnetization. The experimentally observed contrast is consistent with the Lorentz TEM simulations of Bloch-type bubble (Figs. 1s-t) in terms of a multi-slice Fourier approach[23] (see Supporting Information), where the domain wall twist continuously rotates perpendicular to the radial direction (as illustrated in the schematic in Fig. 1r).

From the defocused Lorentz TEM observations in the above two samples, the magnetic contrast at zero tilt is quite different: no contrast for skyrmions in heterostructure FGT (Fig. 1e) versus clear contrast for bubbles in fabricated FGT (Fig. 1l). While there is no doubt on the Bloch-type bubbles in the fabricated FGT, the type of magnetic twist (skyrmion) in the heterostructure FGT is still unclear, even though the Lorentz TEM observations in Figs. 1d-f seem to be consistent with the simulated Néel-type skyrmion (Figs. 1p-q). Néel-type skyrmions, whose magnetic moments are oriented radially (Figs. 1o-p), do not induce contrast-forming electron deflections due to the cancelling-out of azimuthal electron deflections when the sample is not tilted. On the other hand, the invisible contrast in defocused Lorentz TEM images observed at zero tilt angle could also be attributed to the Bloch-type twists where the domain wall is too narrow to be detected by defocused Lorentz

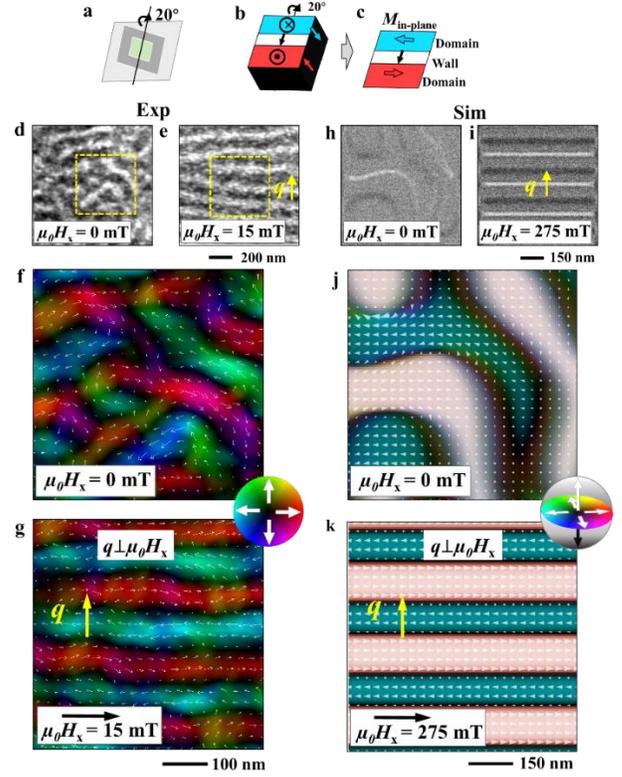

Fig. 2. Néel-type twists with $q \perp \mu_0 H_x$ in a heterostructure FGT. (a) Sample tilt schematic. (b-c) Illustration of the in-plane component projection ($M_{\text{in-plane}}$) from the tilted out-of-plane moments. (d-e) Lorentz TEM images and (f-g) corresponding magnetic inductions resulting from an external in-plane magnetic field $\mu_0 H_x$ applied on the Néel-type dominated twists in the Au-capped heterostructure FGT: (d, f) at $\mu_0 H_x = 0$ mT and (e, g) at $\mu_0 H_x = 15$ mT, revealing an alignment of the modulation $q$-vector perpendicular to the field direction under the in-plane field. (h-i) Multi-slice Lorentz TEM simulations of micromagnetically simulated spin textures and (j-k) corresponding in-plane magnetic induction maps of Néel-type dominated twists with (h, j) $\mu_0 H_x = 0$ mT and (i, k) $\mu_0 H_x = 275$ mT, which qualitatively agrees with (d-g). 'Exp' and 'Sim' stand for 'experiment' and 'simulation', respectively.

TEM measurements, as reported in a centrosymmetric FM[14].

To identify the domain wall twist of skyrmions in the heterostructure FGT, we apply an in-plane field on the sample and use Lorentz TEM to probe the field-aligned magnetic twists. The Lorentz TEM observations were performed at 20° sample tilt (as indicated in Fig. 2a): after ZFC from RT to 143 K, the maze domains show up in Fig. 2d; after in-plane field (15 mT) cooling from RT to 143 K, a single-$q$ state appears with a $q$ vector perpendicular to the field direction (Fig. 2e). The corresponding magnetic induction field maps (Figs. 2f-g and Fig. S4) analyzed from defocused Lorentz TEM images indicate the in-plane projections of the out-of-plane moments at tilt, and not the twisted domain walls (as described by the schematics of Figs. 2b-c).

We have compared the experimental observations of in-plane field-aligned magnetic twists with micromagnetic simulations of Néel-type dominated twists (see Methods) (Figs. 2h-k). The simulated Néel-type maze domains (Fig. 2h, 2j) are relaxed from the paramagnetic state at zero field. While holding an in-plane field of 275 mT to relax the magnetic twists, they are aligned well with a $q$ vector perpendicular to the in-plane field direction. Their projected magnetic induction field maps (Figs. 2j-k) at 20° tilt are calculated from those at 0° tilt. The simulated Lorentz TEM image contrasts of the maze domains (Fig. 2h) and the single-$q$ state (Fig. 2i) coincide with the experimental observations in the heterostructure FGT at 20° (Figs. 2d-e). The consistency of the experimental observations (Fig. 2d-g) and the simulations (Fig. 2h-k) suggests that the observed magnetic twists in the heterostructure FGT (Figs. 1c-f and Figs. 2d-g) are dominantly Néel-type in nature.

In comparison with the Néel-type twists in the heterostructure FGT, we have also applied in-plane fields on the Bloch-type twists in the fabricated FGT (Fig. 3). Lorentz TEM observations are performed at zero tilt since Bloch-type twists impart a measurable phase change on the passing electron wave, resulting in a build-up of intensity. Figures 3a and 3c show the Lorentz TEM images and magnetic induction field maps of the maze domains obtained after cooling to 100 K at 0 mT. The magnetic twists align to a single-$q$ state after FC at $\mu_0 H_x = 20$ mT (Fig. 3b, 3d), where the $q$-vector rotates and aligns at an angle of ∼53° with respect to the in-plane field direction. Note that the aligned twists in Fig. 3d are "fan-like twists" since the domain wall twists are oriented along a single direction, i.e., they tend to be polarized with the in-plane field. The Lorentz TEM image of Fig. 3b shows that the domain-wall contrasts hold alternating half-white and half-dark contrasts (as described by the inserted schematics).

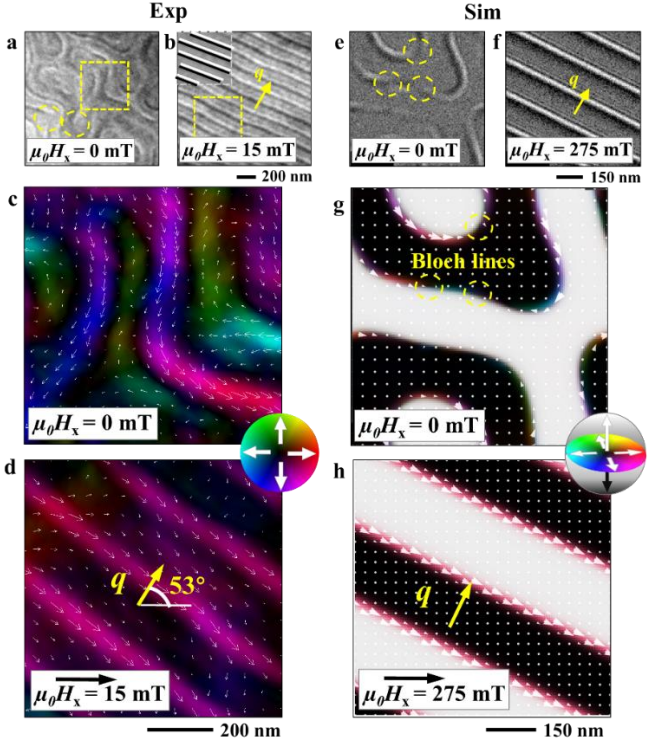

Fig. 3. Bloch-type twists rotated via an external in-plane magnetic field ($\mu_0H_x$) applied on a fabricated FGT thin plate. (a-b) Lorentz TEM images and (c-d) corresponding magnetic inductions at (a, c) $\mu_0H_x = 0$ mT and (b, d) $\mu_0H_x = 15$ mT, reveal an alignment of the modulation $q$-vector at an angle of 53° with respect to the field direction. (e-f) Multi=slice Lorentz TEM simulations applied on micromagnetically simulated spin textures and (g-h) corresponding magnetic in-plane induction maps of Bloch-type twists with (e, g) $\mu_0H_x = 0$ mT and (f, h) $\mu_0H_x = 275$ mT, which qualitatively agrees with (a-d). Bloch lines present in the domain walls are circled in yellow in (a, e, and g). The inserted schematics in (b) describe the black/white contrasts of domain walls in Lorentz TEM images.

We have further performed micromagnetic simulations for Bloch-type twists. The simulated maze domains (Fig. 3e, 3g) are relaxed from a randomly distributed spin configuration at 0 mT. A single-$q$ state with fan-like twists forms upon FC (Figs. 3f, 3h). The simulated Lorentz TEM image contrast, magnetic induction field maps of maze domains and the rotated-$q$ twists (Figure 3e-h) coincide with the experimental observations in fabricated FGT (Figs. 3a-d). Fan-like twists with a rotated-$q$ vector in the present fabricated FGT is slightly different from the field-aligned twists in the chiral magnets exhibiting DMI, where the $q$ aligns parallel to the in-plane field direction. It is possibly because there are many magnetostatic charge-induced Bloch-lines in the present FGT with the dipolar interaction, as marked by yellow circles in Figs. 3a, 3e and 3g and Fig. S1. The Bloch lines are easily tuned to orient to the in-plane field direction, and hence it may facilitate the emergence of fan-like twists.

Our combined experimental and simulation efforts show that the magnetic twists in the heterostructure FGT hold Néel-type dominated nature identified by the $q \perp \mu_0\boldsymbol{H}_x$ alignment under the in-plane field together with the contrast changes in a series of tilted Lorentz TEM images. The identification of Néel-type twists suggests a possible inherent interfacial DMI existing in the heterostructure FGT, intrinsically different from the trivial bubbles stabilized by the magnetic dipolar interaction in the fabricated FGT thin plate. The DMI energy stabilizing the Néel-type twists arising from the asymmetric interface between the Pt layer, oxidized-FGT and FGT itself is predicted in band calculations[9]. The local symmetry breaking of the FGT crystal structure, the magnetic coupling between FGT and Pt capping layer, and/or the asymmetric strain effects (arising from the difference between the free top surface of FGT and its confined bottom surface on the $Si_3N_4$ membrane) may also contribute to the asymmetric interaction. The sample thickness is also an important factor to affect the Néel or Bloch-type twists. A detailed theoretical explanation requires further studies. There is, however, clear experimental evidence coupled with micromagnetic simulations that the Néel-type nature of magnetic twists is dominant in the heterostructure FGT. On the other hand, we have shown that Bloch-type twists in the fabricated FGT thin plate have many intrinsic Bloch-lines (see Fig. 3a, 3e, 3g and Fig. S1) at zero field and orient with a rotated *q* vector with fan-like modulations under in-plane fields. These Bloch lines may facilitate the transition from maze domains into fan-like modulations via a zipper-like mechanism. These characteristics show a remarkable difference from those in the heterostructure FGT (Fig. 1 and Fig. 2) and chiral magnets[19,20].

In summary, we have demonstrated the alignment of magnetic twist *q* vector perpendicular to the in-plane field direction in heterostructure FGT and a rotated *q* vector in the fabricated FGT, which coincides with micromagnetic simulations. Comparing Lorentz TEM observations of magnetic twists in the heterostructure FGT and the fabricated FGT, we conclude here that we may tune Néel-Bloch twists by engineering capping layers in vdW-structure systems. Distinguishing and tuning Bloch- versus Néel-type magnetic twists is a significant aid in researchers' quest to understand the physical nature in magnets and engineer 2D magnetic spintronic devices.

**Acknowledgements**

The authors thank J. Masell for helpful discussion and T. Kikitsu and D. Hashizume (Materials Characterization Support Team in the RIKEN Center for Emergent Matter Science) for technical support on the TEM (JEM-2100F). This work was funded in part by Grants-In-Aid for Scientific Research (A) (Grant No. 19H00660) from JSPS and Japan Science and Technology Agency (JST) CREST program (Grant No. JPMJCR20T1). X.Z. was an International Research Fellow of JSPS. X.Z. was supported by JSPS KAKENHI (Grant No. JP20F20363).

**Conflict of Interest**

The authors declare no conflict of interest.